\documentclass[final,3p,times]{elsarticle}

\usepackage{graphicx}
\usepackage{subfigure}
\usepackage[center]{caption}
\usepackage{listings}
\usepackage{color}
\definecolor{ForestGreen}{RGB}{34,139,34}
\usepackage{amssymb}

\usepackage{amsthm}
\usepackage{amsmath}

\usepackage{fancyvrb}
\DefineVerbatimEnvironment{code}{Verbatim}{fontsize=\small}
\DefineVerbatimEnvironment{example}{Verbatim}{fontsize=\small}

\begin{document}

\begin{frontmatter}


\author{E.I. Ioannidis}
\author{N. Cheimarios\corref{cor1}}
 \ead{nixeimar@chemeng.ntua.gr}
 \cortext[cor1]{Tel.: +30 210 772 3296 \ Fax: +30 210 772 32 97}

\author{A.N. Spyropoulos}
\author{A.G. Boudouvis}

\title{On the performance of various parallel GMRES implementations on CPU and GPU clusters}

\address{School of Chemical Engineering, National Technical University of Athens, Athens 15780, Greece}
\begin{abstract}
As the need for computational power and efficiency rises, parallel systems become increasingly popular among various scientific fields. While multiple core-based architectures have been the center of attention for many years, the rapid development of general purposes GPU-based architectures takes high performance computing to the next level. In this work, different implementations of a parallel version of the preconditioned GMRES - an established iterative solver for large and sparse linear equation sets - are presented, each of them on different computing architectures: From distributed and shared memory core-based to GPU-based architectures. The computational experiments emanate from the dicretization of a benchmark boundary value problem with the finite element method. Major advantages and drawbacks of the various implementations are addressed in terms of  parallel speedup, execution time and memory issues. Among others, comparison of the results in the different architectures, show the high potentials of GPU-based architectures.

\end{abstract}


\end{frontmatter}



\section{Introduction}
\label{intro}

Applying finite element methods on boundary value problems involves the solution of large and sparse systems of linear (or linearized) algebraic equations of the form 
\begin{equation}
\label{original}
{\bf Ax} = {\bf b}
\end{equation}
an expensive computational task that requires most of the total processing time.  To accelerate the computations, parallel processing in conjunction with domain decomposition methods \cite{Smith} (DDM) are commonly used. DDM are separated in three major categories: Schwarz-like, Schur complement and full matrix methods. Applying Schwarz-like or Schur complement methods amounts to splitting the original system into smaller ones that are solved separately and in parallel
while full matrix methods solve in parallel the original system (Eq.\ref{original}). Elaborating on the advantages and disadvantages of the various DDM is outside the scope of this paper but the interested reader is referred to \cite{Saad_book} (Chapter 13) and references therein. 

In this paper, the parallel implementation is based on a Krylov-type solver that iterates on the original system (full matrix based DDM). Krylov-type iterative solvers are commonly used for such large and sparse systems due to their small memory requirements and high parallel efficiency. The GMRES (Generalized Minimal Residual) method \cite{Saad} is applied here; an iterative solver commonly used for general nonsymmetric matrices that is based on efficiently parallelizable operations like matrix-vector products and vector inner products. Furthermore, to enhance the convergence of the method, a preconditioning technique by deflation is employed in our computational experiments. Henceforth, the full matrix DDM implemented for the parallel solution of a finite element problem along with the preconditioned GMRES solver is referred to as \emph{parallel GMRES} (PGMRES). 

Concerning the hardware, distributed, shared and hybrid memory systems were until recently the main computing architectures used for parallel computations. Sequential algorithms were transformed to run in those parallel architectures exploiting their high computing capabilities. Nevertheless, the constant need for solving efficiently demanding problems has led to the development of a new kind of parallel computing systems. Driven by the demands of the gaming industry, graphics processing units (GPUs) have evolved over the years offering remarkable floating point arithmetic performance \cite{papadra}, often exceeding 1 TFlops for a single GPU device. Their use for general purpose computations is known as general purpose computing on GPUs (GPGPU) and has become extremely popular over the last few years. With the release of the first NVIDIA's Computing Unified Device Architecture (CUDA\cite{ncuda}) Software Development Kit (SDK) in 2007, the tricky GPGPU programming became relatively easy. CUDA is a parallel computing framework developed by NVIDIA, which gives developers the capability to directly access GPU's memory and instruction set.

The past few years have seen intensive research on numerical methods and parallel processing. Efficient parallel implementations of the GMRES on distributed memory core-based clusters are reported by Haiwu \cite{GMRES_distr}, Li \cite{Li}, Erhel \cite{Erhel} and Pashos et al. \cite{Pashos1}, while Calvetti et al. \cite{Calvetti} implemented on shared memory architectures. In the field of GPGPU matrix computations, Galoppo et al. \cite{Galoppo} studied the peak performance of a GPU dense matrix decomposition algorithm. Furthermore, Volkov et al. \cite{Volkov} improved the standard CUBLAS algorithm for matrix-matrix multiplication, claiming to approach peak performance on a G80 series NVIDIA GPU. Sparse matrix algorithms on manycore GPUs were presented by Garland \cite{Garland}, while Boltz et al. \cite{Boltz} were among the first to report an implementation of a GPU-based iterative solver. These early works generally focused on tackling the various problems that arised in early numerical computations on GPUs and underlined the high computing capabilities and potential that these systems have. Focusing on the GMRES, several works like  \cite{sparse_GPU} - \cite{Wang} implement GMRES in GPU-based computing architectures, proposing efficient \textsl{kernels} and special storage formats for accelerating the computations several times. Furthermore, a hybrid GMRES implementation is presented by Moret \cite{Moret}, based on GRID systems, while multi-GPU implementations of various numerical methods, like those presented in \cite{Xian} and \cite{Ament}, show promising results regarding the parallel computing capabilities of multi-GPU systems. 

Despite the extensive literature around PGMRES solver, there is a significant shortage of comparisons between various implementations of PGMRES on different computing  architectures (distributed, shared and GPUs). This article focuses on illuminating the implementation of a single, custom parallel finite element method with the PGMRES in various computing architectures. The main goal is to compare the major advantages and drawbacks of each parallel computing implementation as far as parallel speedup, execution time and memory issues are concerned. That said, three codes are developed:

\begin{enumerate}
\item PGMRES-classic: PGMRES capable of executing in multi-processing units. The data passing is performed with Message Passing Interface (MPI).
\item PGMRES-GPU: A single GPU-based PGMRES. The parallel tasks (communication, synchronisation) are performed via CUDA. 
\item PGMRES-mGPU: A multiple GPU-based PGMRES. The parallel tasks (communication, synchronisation) in each GPU and between the GPUs are performed via CUDA.
\end{enumerate}

This paper begins with an introduction to the partitioning technique (Section \ref{partitioning}) used here for the distribution of computational work between the various processing units. Section \ref{hardware} describes the architectures used in the experiments, giving specific details about the hardware available for computations. The Bratu benchmark problem is briefly described in Section \ref{bratu} followed by a short presentation of the preconditioned GMRES in Sections \ref{GMRES} and \ref{DEFLGMRES}, while experimental results and comparison of the various architectures are shown in Section \ref{results}. Finally, conclusions are presented in Section \ref{conclusions} and further research directions are proposed. 

\section{Computing}
\label{computing}

\subsection{Partitioning}
\label{partitioning}

In order to apply PGMRES, a graph partitioning technique for the original matrix is required \cite{metis}. Graph partitioning aims at the optimal distribution of the computational load between the multiple cores towards high computational speed. For an efficient partitioning, a balanced load is critical otherwise it will lead to a performance bottleneck, caused by the most heavy loaded core. Furthermore, solving the partitioned problem requires communication and synchronisation between the cores. High and frequent data exchange load between the cores has negative effect on the parallel speedup and can make parallel processing inefficient compared to sequential execution of a program. Thus, partitioning techniques should minimize this effect. The graph partitioning corresponds to the finite element mesh partitioning, i.e. splitting the original domain into several subdomains each assigned to a single core. Each subdomain is processed independently, assembling a corresponding local matrix and residual vector. In order to reduce the effects of the choice of the partitioning technique, i.e. the "quality" of the produced subdomains, a simple, 3D, cubic domain is used in the present computations. At this point we should point out that the communication required by the various processes can be described as local, if it occurs between processes sharing neighboring subdomains or as global, if it occurs between all concurrently running processes.

\subsection{Hardware}
\label{hardware}
 In this paper, a SPMD (single program, multiple data) technique is used for parallel processing involving both cores and GPUs. Shared memory and distributed memory systems are both included in the computational experiments. All computations are carried out in two clusters, hosted by the School of Chemical Engineering of the National Technical University of Athens, Pegasus and Andromeda. The cluster Pegasus \cite{h_pegasus} is a distributed memory system and consists of 16 nodes interconnected via Myrinet-2000 network. Each of these nodes comprises 2 Xeon CPUs at 3 GHz with 2GB RAM for a total of 32 CPUs and 32GB RAM. The cluster Andromeda \cite{h_andromeda} is a distributed memory hybrid (CPU/GPU) cluster. It includes 4 nodes (HP ProLiant SL390s G7) each consisting of 2 Xeon 6-core CPUs X5660 at 2.80 GHz with 16GB RAM for a total of 48 cores and 64GB RAM connected via Gigabit Ethernet network. Furthermore 2 NVIDIA GPUs Tesla M2050 with 3GB RAM are on each node for a total of 8 GPUs and 24GB RAM. Each NVIDIA GPU hosts 14 multiprocessors, each containing 32 GPU-cores for a total of 448 GPU-cores. In this work, only one node (shared memory architecture) of Andromeda is used in the experiments.  

\section{Benchmark problem and GMRES}
\subsection{The Bratu problem}
\label{bratu}

The benchmark problem employed here is the single parameter, 3-D Bratu problem (\ref{bratu-eq}) - (\ref{bratu-bound-z}). This boundary value problem was posed in a 3-D cubic domain in cartesian coordinates with homogeneous Dirichlet boundary conditions on the $x-$ and $y-$boundaries and homogeneous Neumann boundary conditions on the $z-$boundaries:

\begin{equation}
\label{bratu-eq}
\nabla^2u+\lambda e^u=0 \hspace{1cm} (x,y,z)\in \Omega,  \hspace{0.3cm} \Omega=[0,1] \ x \ [0,1] \ x \ [0,1]
\end{equation}
\begin{equation}
\label{bratu-bound-x}
u|_{x=0}=0 \hspace{0.1cm} \hspace{0.3cm} u|_{x=1}=0 \hspace{0.2cm} \forall \ (y, z) \in \Omega
\end{equation}
\begin{equation}
\label{bratu-bound-y}
u|_{y=0}=0 \hspace{0.1cm} \hspace{0.3cm} u|_{y=1}=0 \hspace{0.2cm} \forall \ (x, z) \in \Omega
\end{equation}
\begin{equation}
\label{bratu-bound-z}
\frac{\partial u}{\partial z}\big\vert_{z=0}=0 \hspace{0.1cm} \hspace{0.3cm} \frac{\partial u}{\partial z}\big\vert_{z=1}=0 \hspace{0.2cm} \forall \ (x, y) \in \Omega
\end{equation}

In the literature \cite{Pashos2}, the 3-D Bratu problem is often studied in terms of its solution space structure and, in particular, the dependence of singularities on parameter values (like $\lambda$). In this implementation though, singularities are not of interest and a constant value $\lambda=6.8$ is chosen.
The above problem is discretized and solved with the Galerkin/finite element method \cite{fem}. The domain is divided into a mesh made of a finite number of 27-node cubic elements. The unknown function u is approximated in terms of finite element basis functions, $\phi_j$, that are quadratic polynomials in each spatial direction:
\begin{equation}
\label{approx}
u(x,y,z)=\sum_{j=1}^N u_j \phi _j(x,y,z)
\end{equation}
where $N$ is the total number of nodes on the mesh covering $\Omega$ and $u_j$ are the values of the unknown function u$(x,y,z)$ at the nodes. Each basis function, $\phi_j$ is so defined as to be equal to unity at node $j$ and zero at all other nodes. The Galerkin method forms $N$ weighted residuals by multiplying equation (\ref{bratu-eq}) by the same basis functions used to approximate the solution $u$ (cf. Eq. (\ref{approx})) and then integrating over the domain $\Omega$:
\begin{equation}
\label{residuals}
R_i=\int_{\Omega}\nabla^2u \phi _i dV+\int_{\Omega}\lambda e^u \phi _idV, \hspace{0.2cm} \ i=1,2,\dots,N
\end{equation}
Following substitution from Eq. (\ref{approx}), Eq. (\ref{residuals}) is transformed into a system of $N$ nonlinear algebraic equations:
\begin{equation}
\label{weak-form}
R_i=\oint_{\partial\Omega} \phi_i \textbf{n} \cdot \nabla u \ dA - \sum_{j=1}^N u_j \int_{\Omega}\nabla \phi_j \cdot \nabla\phi_i \ dV+\lambda \int_{\Omega}exp({\sum_{j=1}^N u_j \phi_j})\phi_i \ dV=0, \hspace{0.2cm} \ i=1,2,\dots,N
\end{equation}
where \textbf{n} is the unit normal vector to the boundary $\partial \Omega$.
Due to the boundary conditions (\ref{bratu-bound-x}) and (\ref{bratu-bound-y}), all the residual equations (\ref{weak-form}) corresponding to nodes on the $x$-faces and $y-$faces of $\Omega$ are replaced by equations setting the value of the local nodal unknowns equal to $0$. Moreover, the boundary integral term in equations (\ref{weak-form}) vanishes in the residuals of all interior nodes of the domain since, by their definition, the basis functions corresponding to interior nodes are zero at the domain boundary. Due to boundary condition (\ref{bratu-bound-z}), the boundary integral term for the nodes corresponding to the $z-$faces of the problem domain vanishes as well. Therefore, there is no boundary integral contribution in the residual equations (\ref{weak-form}). The nonlinear equation set (\ref{weak-form}) is linearized and solved by Newton iteration. At the $k-th$ iteration, the following linear equation set is solved:

\begin{equation}
\label{system}
\mathbf{J}(\mathbf{u}^k,\lambda)\mathbf{du}^k=-\mathbf{R}(\mathbf{u}^k,\lambda)
\end{equation}
where $\mathbf{du}$ is the update vector; $\mathbf{u}^{k+1}=\mathbf{u}^k+\mathbf{du}^k$. $\mathbf{R}=[R_i]$, $i=1,2,\dots,N$, is the residuals vector; $R_i$ is given by Eq. (\ref{weak-form}) with the boundary integral term eliminated. $\mathbf{J}=[J_{ij}]$, $i, j=1,2,\dots,N$, is the Jacobian matrix; $J_{ij}\equiv \frac{\partial R_i}{\partial u_j}$. The superscript k in Eq. (\ref{system}) denotes evaluation at the current approximation of the solution $\mathbf{u}$.

The Newton iteration terminates when a selected norm of the update vector $\mathbf{du}^k$ falls below a tolerance. The linear system (\ref{system}) is solved, at each iteration by the preconditioned GMRES with restarting, GMRES(m) \cite{Saad} (see below). 

\textit{Henceforth the term "degrees of freedom - DOF" is used to refer to the unknowns the equations are solved for. Also, the matrix \textbf{J}, the right hand side -\textbf{R} and the solution vector \textbf{du} in Eq. (\ref{system}) will be denoted, respectively, by \textbf{A}, \textbf{b} and \textbf{x}, to conform with the established notation (used in Eq. (\ref{original}))}. 

\subsection{GMRES(m)}
\label{GMRES}
Starting from an initial guess, $\mathbf{x_0}$, of the solution of (\ref{system}), GMRES employs Arnoldi method, combined with an orthogonalization technique - the modified Gram-Schmidt method is used here - to construct an orthonormal basis $\mathbf{V_m} \in \mathbf{R}^{Nxm}$ of the m-dimensional Krylov subspace as follows

\begin{equation}
\label{krylov-basis}
K_m(\mathbf{A},\mathbf{\upsilon})=span\left\lbrace \mathbf{\upsilon},\mathbf{A}\mathbf{\upsilon},\mathbf{A}^2\mathbf{\upsilon},\dots,\mathbf{A}^{m-1}\mathbf{\upsilon} \right\rbrace
\end{equation}
where $\mathbf{\upsilon} \equiv \frac{\mathbf{r_0}}{\|\mathbf{r_0} \|_2}$ and $\mathbf{r_0}\equiv \mathbf{b}-\mathbf{A}\mathbf{x_0}$ the GMRES residual vector. The new approximation of the solution is computed as:

\begin{equation}
\label{new_sol}
\mathbf{x_m}=\mathbf{x_0}+\mathbf{V_m}\mathbf{y_m}
\end{equation}
where $\mathbf{y_m}$ is a vector of size $m$, which comes out of the least squares problem

\begin{equation}
\label{arg}
\mathbf{y_m}=\underset{\mathbf{y}}{\operatorname{argmin}} \| \beta \mathbf{e_1} - \mathbf{\bar{H}_m} \mathbf{y_m} \|_2.
\end{equation}
In Eq. (\ref{arg}), $\beta \equiv \|\mathbf{r_0}\|_2$, $\mathbf{e_1}=[1, 0, \dots, 0]^T$ and $ \mathbf{\bar{H}_m} \in \mathbf{R}^{(m+1)xm}$ is an upper Hessenberg matrix that satisfies the equation

\begin{equation}
\label{hessenberg}
\mathbf{A}\mathbf{V_m}=\mathbf{V_{m+1}}\mathbf{\bar{H}_m} \Rightarrow \mathbf{V_m^T}\mathbf{A}\mathbf{V_m}=\mathbf{H_m}
\end{equation}
where $\mathbf{H_m} \in \mathbf{R}^{m x m}$ is an upper Hessenberg matrix obtained from the $\mathbf{\bar{H}_m}$ by deleting its last row. The least squares problem (\ref{arg}) is solved by transforming $\mathbf{\bar{H}_m}$ into an upper triangular matrix $\mathbf{R_m} \in \mathbf{R}^{mxm}$ using plane rotations. 

The residual vector at every $k$ step ($k$ = 1, 2 . . ., $m$) of the GMRES is 
\begin{equation}
\label{res}
\|b - Ax_k\|_2 = |\gamma_{k+1}| 
\end{equation}
where $\gamma_{k+1} \in R$ is a by product of the $\mathbf{\bar{H}_m}$ plain rotations. Thus, at every $k$ step, the convergence of the GMRES can be monitored from Eq.(\ref{res}) without computing the $\mathbf{x_m}$.

GMRES allows the Krylov subspace dimension to increase up to $N$ and always terminates in at most $N$ iterations. Since the GMRES 
requires $\mathcal{O}(mN)$ memory storage and $\mathcal{O}(mN^2)$ floating point operations, for larger values of m it becomes very expensive. Therefore a restarting variant of the GMRES - the GMRES(m) \cite{Saad} - is used in practise. When $m$ reaches a certain preset value, the algorithm restarts, using the last approximation $\mathbf{x_m}$ from (\ref{new_sol}) as a new initial guess \cite{Pashos1}. Thus, two iterations are performed: The "inner" $m$ iterations and the "outer" iterations that correspond to the restarts of the GMRES(m). The termination criteria is the relative error $\varepsilon$ defined as
\begin{equation}
\label{relative}
\varepsilon =\frac{\| \mathbf{r}\| _2}{\| \mathbf{r_0} \| _2}
\end{equation}
to fall below a tolerance. In Eq. (\ref{relative}), $\|\mathbf{r}\|_2$ is the Euclidean norm of the residual vector $\mathbf{r}=\mathbf{b}-\mathbf{A} \mathbf{x_m}$ after the last outer iteration of GMRES(m) and $\|\mathbf{r_0}\|_2$ the Euclidean norm of the initial residual vector $\mathbf{r_0}=\mathbf{b}-\mathbf{A} \mathbf{x_0}$. The initial guess $\mathbf{x_0}$ is usually taken equal to zero.

\subsection{Preconditioning by deflation}
\label{DEFLGMRES}

The convergence rate of GMRES depends on the smallest eigenvalues of the Jacobian matrix in Eq.(\ref{system}) and deteriorates as the eigenvalues tend to zero. Moreover, the minimum and maximum eigenvalues of the Jacobian can be approximated by the corresponding eigenvalues of the Hessenberg matrix in Eq.(\ref{hessenberg}). These eigenvalues are known as the Ritz values \cite{mat_computations}. However, the restarted GMRES(m) loses information that is related to the smallest Ritz values at each restart and several preconditioning techniques have been developed that aim at preserving those information through the process \cite{Moret}\cite{Morgan}.

A preconditioner is used to improve the convergence properties of the GMRES(m), especially near singular points, where convergence is difficult and sometimes even impossible. Using the preconditioner, the original linear system is transformed to an equivalent one that has better convergence properties. In our implementation, (\ref{system}) is preconditioned from the right as

\begin{equation}
\label{precond1}
\mathbf{AM^{-1}z}=\mathbf{b}, \ \mathbf{x}=\mathbf{M^{-1}z}.
\end{equation}

In (\ref{precond1}), $\mathbf{z}$ is a vector of size $N$ and $\mathbf{M^{-1}} \in \mathbf{R}^{NxN}$ is the preconditioner matrix which is constructed from a deflation technique \cite{burrage}\cite{erhel} and is given from

\begin{equation}
\label{precond2}
\mathbf{M^{-1}}=\mathbf{I_N}+\mathbf{U}\left(|\mu |\mathbf{T_r^{-1}}-\mathbf{I_r}\right)\mathbf{U^T}
\end{equation}

where $\mu \in \mathbf{R}$ is the largest eigenvalue of the Jacobian matrix, $\mathbf{I_N} \in \mathbf{R}^{NxN}, \ \mathbf{I_r} \in \mathbf{R}^{rxr}$ are identity matrices, $\mathbf{U} \in \mathbf{R}^{Nxr}$ is an orthonormal basis of the $r$-dimensional invariant subspace, $P_r$, corresponding to the $r$ smallest (in absolute value) eigenvalues of the Jacobian and $\mathbf{T_r} \in \mathbf{R}^{rxr}$ such as

\begin{equation}
\label{tau_mat}
\mathbf{T_r}=\mathbf{U^TAU}.
\end{equation} 

The largest eigenvalue needed in Eq. (\ref{precond2}), and the eigenvectors of the Jacobian needed in Eq. (\ref{tau_mat}), are approximated by those of the Hessenberg matrix $\mathbf{H_m}$. 
At each restart of the GMRES(m), $l$ eigenvectors of the Jacobian are approximated from $l$ eigenvectors, $\mathbf{z_i}\in \mathbf{R^{rxr}}$ of $\mathbf{H_m}$ that correspond to the smallest eigenvalues. These eigenvectors are computed using the inverse power method. In practice, we use $l=1$. The new vectors: 

\begin{equation}
\label{precond3}
\mathbf{u_i}=\mathbf{V_m z_i}, \ i=1,\dots,l
\end{equation}

are orthonormalized against those of $\mathbf{U}$ and added to $\mathbf{U}$. So the dimension $r$ of $P_r$ increases by $l$. 
In practice, the dimension $m$ of $\mathbf{H_m}$ is small, therefore the computational cost of the eigenvalue problem for $\mathbf{H_m}$ is negligible. 

The preconditioner needs two arrays of size $2Nr$ for the storage of $\mathbf{U}$ and $\mathbf{AU}$ involved in Eq. (\ref{precond2}) and (\ref{tau_mat}). In order to reduce the memory requirements and computational cost, an upper limit $r_{max}$ on $r$ is set ($r_{max}=20$ in our case). When $r>r_{max}$ the update of the preconditioner is continued (i.e. $l$ new vectors are added to $P_r$) but at the same time, $l$ vectors of $P_r$ that correspond to the $l$ largest eigenvalues of the matrix $\mathbf{T_r}$ are deflated.

The orthonormal basis $\mathbf{U}$ of $P_r$ is constructed from the eigenvectors $\mathbf{z_i}$ of $\mathbf{H_m}$ times the orthonormal basis $\mathbf{V_m}$ of the Krylov subspace. Thus, the vectors $\mathbf{u_i}$ approximate the eigenvectors of the Jacobian matrix that correpsond to the smallest eigenvalues. 

\section{Results}
\label{results}
In this section, the results obtained from four different computational experiments are presented: \\

\noindent {\bf Case A}: PGMRES-classic on a distributed memory architecture (Pegasus cluster). A maximum number of twelve single core processors (one per node) from Pegasus cluster is used. \\
\noindent {\bf Case B}: PGMRES-classic on a shared memory architecture (Andromeda cluster). A maximum number of twelve cores belongs to each Andromeda node. \\
\noindent {\bf Case C}: PGMRES-GPU on a single GPU (one node of Andromeda cluster) \\
\noindent {\bf Case D}: PGMRES-mGPU on two GPUs (one node of Andromeda cluster) \\

\noindent The main quantity monitored during the experiments is the \textsl{speedup} performance achieved from the implementation of PGMRES on various architectures. Furthermore, the effect of the required communication time on parallel implementations is analyzed. The findings are strongly dependent on the size of the problem and therefore, all implementations are tested in a wide range of sizes, starting from a few thousands DOF up to a million.

PGMRES-classic is written in C with MPI and compiled with Intel C/C++ standard \texttt{icc} compiler, without any compilation flags, combined with OpenMPI library for communication in Andromeda and with mpich-gm library for communication in Pegasus. PGMRES-GPU and PGMRES-mGPU are written in C with CUDA. The C code is compiled with Intel C/C++ standard \texttt{icc} compiler while for the GPU codes, NVIDIA's Cuda standard \texttt{nvcc} compiler  is used. No compilation flags are used. All computations are made with double precision floating point data for real variables.  

In all PGMRES versions, a \textsl{Compressed Sparse Row (CSR or CRS)} format is adopted for storage of large sparce matrices arising from the finite element problem. The following table contains information about the different size, sparsity pattern, total number of entries and memory requirements of the compressed matrix \textbf{A}, as the number of DOF grows from some thousands to over a million.

\begin{table}[h!]
\begin{center}
\caption{Sparsity pattern of matrix \textbf{A} }
\begin{tabular}{ccccc}
\hline\noalign{\smallskip}
DOF & Matrix elements & Nonzero entries & Sparsity & Memory (MBytes)\\ 
\noalign{\smallskip}\hline\noalign{\smallskip}
29791 & 8.88E+08 & 1.60E+06 & 1.80E-03 & 19  \\
132651 & 1.76E+10 & 7.72E+06 & 4.39E-04 & 88\\
226981 & 5.15E+10 & 1.33E+07 & 2.58E-04 & 153 \\
357911 & 1.28E+11 & 2.13E+07 & 1.66E-04 & 245 \\
531441 & 2.82E+11 & 3.06E+07 & 1.08E-04 & 380 \\
753571 & 5.68E+11 & 4.55E+07 & 8.01E-05 & 524 \\
1030301 & 1.06E+12 & 6.26E+07 & 5.89E-05 & 720 \\
1367631 & 1.87E+12 & 8.35E+07 & 4.46E-05 & 961 \\ 
1771561 & 3.14E+12 & 1.06E+08 & 3.46E-05 & 1249 \\ 
\noalign{\smallskip}\hline
\end{tabular}
\label{sparsity_t}
\end{center}
\end{table}

\begin{figure}[h!]
\centering
  \includegraphics[scale=0.95]{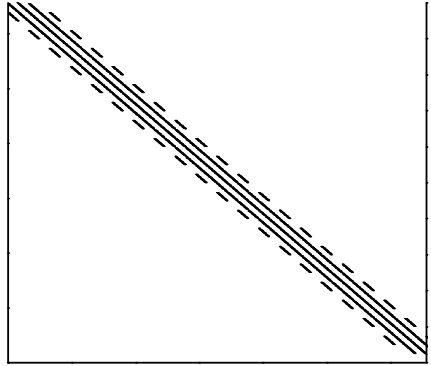}
\caption{Sparsity pattern of matrix \textbf{A}}
\label{sparsity}       
\end{figure}

As far as the GMRES(m) is concerned, in all experiments the number of both inner ($m$) and outer iterations is kept constant and equal to 50 and 100 respectively. The number of outer iterations, in particular, is taken constant in order to keep the computational results unaffected by accuracy variations and roundoff errors that may occur. Therefore, no termination criteria are used. Nevertheless, relative errors (see Eq. (\ref{relative})) remain for all problems below $10^{-3}$.

\subsection{GMRES(m) vs preconditioned GMRES(m)}

\begin{figure}[h!]
\centering
  \includegraphics[scale=0.9]{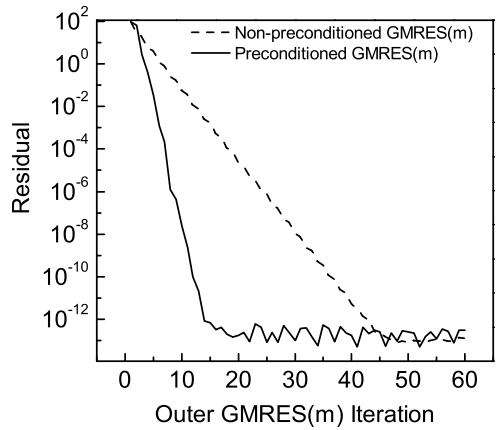}
\caption{Convergence of GMRES(m) with and without preconditioner}
\label{converge}       
\end{figure}

A preconditioner is used to enhance the convergence behavior of the GMRES. To elucidate the effect of the preconditioner, Figure \ref{converge} shows the Euclidean norm of the residual vector ($\mathbf{r}=\mathbf{b}-\mathbf{A} \mathbf{x_m}$) during the solution of a system with DOF=130000 versus the outer iterations of GMRES(m) with and without preconditioner. Figure \ref{converge} clearly depicts the acceleration in the convergence of the preconditioned version versus the non-preconditioned one.

Furthermore, since the preconditioning involves mainly inner products and matrix-vector products (see Eq. (\ref{precond1})-(\ref{precond3})) like the GMRES(m) the overall speedup measured is not affected. Thus, only the results from the preconditioned version of restarted GMRES(m) follow.

\subsection{Case A: PGMRES-classic on distributed memory architecture}

\begin{figure}[h!]
\centering
  \includegraphics[width=0.5\textwidth]{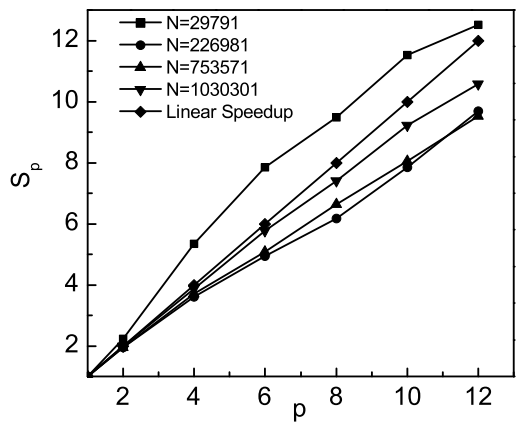}
\caption{Speedup on a distributed memory parallel processing architecture}
\label{pegasus}       
\end{figure}

In Fig.\ref{pegasus} the speedup achieved is shown for various DOF. This speedup \textsl{$(S_p)$} is defined as the serial execution time, \textsl{$T_1$},  (one MPI process), divided by the run time, \textsl{$T_p$}, required by \textsl{p} (in number) MPI processes, for the same DOF,
 \begin{equation}
\label{speedup_core}
 S_p = \frac{T_1}{T_p}
\end{equation}  

Increasing the number of processors leads to a smaller number of computations executed by each MPI process. As a result, the computing time required by multiple processors is lower than the time required by the sequential solution from a single processor. The ideal would be that the speedup and the number of cores relate linearly, so for algorithms with order of growth $\mathcal{O}(n)$, like the GMRES, a speedup of \textsl{$S_{p-ideal}=p$} is expected. On the contrary, in Fig.\ref{pegasus} a super-linear speedup is achieved, occurring for rather small problems, where the acceleration surpasses \textsl{p}. This observation contradicts the analysis made about the correlation between the speedup and the size of the problem. The super-linear speedup is due to \textsl{cache-effects} \cite{Pashos1}. Every processor has a L2 cache with capacity of 2MB. Therefore, problems with less than 30000 DOF assemble a matrix with size of less than 20 MB (see Table \ref{sparsity_t}). The partitioning technique splits the data into chunks of memory smaller than 2 MB that lie permanently in the fast cache memory and not in the slower main memory (RAM) as in the larger problems (DOF $\geq 30000$).

The speedup is, in general, affected by the capabilities of the network used for the communication. In the Pegasus cluster - where the Myrinet network is used - communication time is negligible as it can be seen in Table \ref{pegasus-time} in which the percentage of the communication time for twelve MPI processes vs DOF is presented. Recall (see section \ref{partitioning}) that the existing communication types are two, one at local level and the other at global level.

\begin{table}[h!]
\begin{center}
\caption{Distribution (\%) of computational time versus DOF (twelve MPI processes)}
\begin{tabular}{lcccc}
\hline\noalign{\smallskip}
DOF & Computation \% & Local Communication \% & Global Communication \% & Total Communication \% \\
\noalign{\smallskip}\hline\noalign{\smallskip}
29791 & 71.85 & 9.61 & 18.54 & 28.15 \\
132651 & 76.83 & 3.99 &  19.18 & 23.17 \\
226981 & 92.23 & 2.41 & 	5.37 & 	7.77 \\
357911 & 92.14 & 2.98 & 4.88 &	7.86 \\
531441 & 94.81 & 1.89 & 3.30	 & 5.19 \\
753571 & 95.13 & 1.75 & 	3.12 & 	4.87 \\
1030301 & 96.15 & 1.21 & 2.64 & 	3.85 \\
1367631 & 96.52 & 1.22 & 2.26 & 3.48 \\ 
1771561 & 97.14 & 1.13 & 1.73 & 2.86\\
\noalign{\smallskip}\hline
\end{tabular}
\label{pegasus-time}
\end{center}
\end{table}

\subsection{Case B: PGMRES-classic on shared memory architecture}

\begin{figure}[h!]
\centering
  \includegraphics[width=0.5\textwidth]{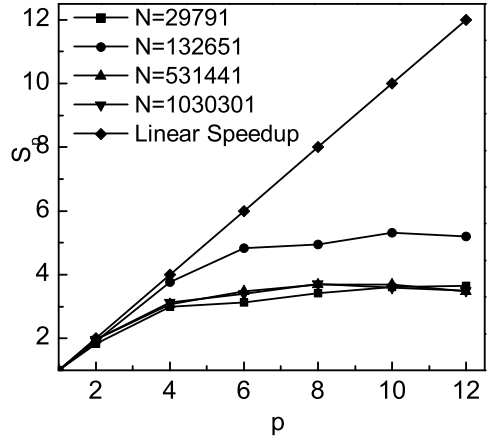}
\caption{Speedup on a shared memory parallel processing architecture}
\label{andromeda}       
\end{figure}

In Fig.\ref{andromeda} the speedup achieved is shown at various DOF. The speedup behaviour of the shared memory architecture changes radically as the number of cores increases. The most noticeable aspect of Fig.\ref{andromeda} though, is the fact that the speedup remains actually constant and lower than the ideal for a number of cores greater than six and for all the experiments compared to the previous distributed architecture where the speedup had a linear variation in the range of two to twelve MPI processes and large (DOF \textgreater 30000) problems. This main difference can be explained by the limitations that the shared main memory puts in this kind of architectures. As the number of parallel executing cores increases, memory accesses become sequential and the cores remain idle, waiting for the data to be fetched. Consequently, core to memory connection becomes a bottleneck and the high computing power of the multi-core system is only partially exploited.

\subsection{Case C: PGMRES-GPU}

No partitioning is used in this implementation, because all instructions are executed concurrently by numerous threads with direct main memory access (see Section. For maximum performance, each instruction must be assigned to a single GPU thread that runs concurrently with the others. Therefore, as the size of the problem grows, so does the number of GPU threads used for computations. The GPU threads are organised in blocks that are executed in parallel by a single GPU multiprocessor. The number of threads per block is pre-defined by the user. The choice here is 512 threads per block. The number of blocks depends on the number of DOF of the problem and is calculated each time by: 
\begin{equation}
	Blocks=\frac{DOF}{512}	
\end{equation}

\noindent After one core finishes the assembly of the \textbf{A} and \textbf{b} in Eq. (\ref{original}), they are uploaded to the GPU where the computations take place. In this case the speedup, $S_{GPU}$ is defined as:
\begin{equation}
S_{GPU} = \frac{T_{1\_core}}{T_{GPU}}
\end{equation}
where \textsl{$T_{1\_core}$} is the run time in one core of Andromeda and \textsl{$T_{GPU}$} is the run time in a single Tesla M2050 GPU.

\begin{figure}[h!]
\centering
  \includegraphics[width=0.5\textwidth]{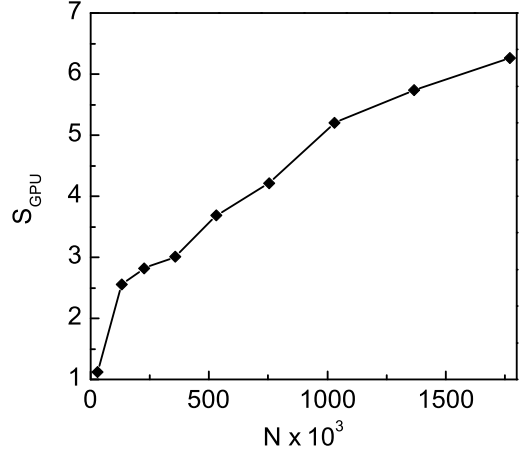}
\caption{Speedup on a single GPU versus problem size}
\label{single-GPU}       
\end{figure}

As shown in Fig.\ref{single-GPU}, the speedup obtained from a single GPU versus DOF shows varies. Firstly, where the number of DOF remains relatively small, the speedup increases sharply and linearly. This trend is explained by the massive parallelism of the GPU with thousands of threads executing concurrently instructions on different segments of data, accelerating the computations as the problem size grows until $10^5$. For DOF \textgreater $10^5$ though, this behavior changes a bit; the speedup increases further but this time at a slower rate. The maximum speedup obtained from a single GPU is 6.5, greater than the maximum speedup measured in the shared memory, but still lower than the distributed memory model's.

Furthermore, another version of the single-GPU implementation was tested. In this version, the partial sums from each dot product and norm computed by the multiple blocks were transferred to the host, followed by the reduction. The final result, computed by the host, was uploaded afterwards back to the GPU. In addition, the least squares problem (Eq. (\ref{arg})) was solved on the host. This implementation involving reduction and synchronisation of the GPU threads by the host is less efficient than the initial single-GPU one. The maximum speedup observed in this version was 4.6, far smaller than the original one.

\subsection{Case D: PGMRES-mGPU}
\label{mGPUs}

With the introduction of CUDA Toolkit 4.0 \cite{ncuda2}, the use of multiple GPUs for general purpose GPU computations became relatively easy. Before 4.0, each GPU had to be controlled by a single host thread, thus making the use of multiple GPUs a complicated and usually inefficient task. The main drawback was the communication between the GPUs requiring data transfers and synchronization between the two core threads. After the introduction of the new Toolkit, multiple GPUs can be controlled by a single core thread making synchronization and communication relatively easy and inexpensive. Furthermore, kernel execution is non-blocking and control is returned to the host before the kernel finishes computing. Therefore multiple kernels can be executed concurrently on multiple devices. The CUDA function \texttt{CudaSetDevice} enables the switching between available GPUs.

In this approach the mesh is partitioned in the same way as in the PGMRES-classic implementation. Each of the two resulting subdomains is assigned to a single GPU but here the data passing is implemented with CUDA instead of MPI. Two speedups are measured, $S_{mGPU}$ and  $S_{g}$ defined, respectively, as the execution time, \textsl{$T_{core}$}, required by a single core of the Andromeda cluster divided by the execution time, \textsl{$T_{mGPU}$}, required by two Tesla M2050 GPUs and as the execution time, \textsl{$T_{GPU}$}, required by a single GPU divided by the execution time, \textsl{$T_{mGPU}$}, required by the multi-GPU implementation. The results for the speedup versus DOF are presented in Figs.2a and b.
\vspace{0.2cm}
\begin{equation} 
S_{mGPU} = \frac{T_{mGPU}}{T_{core}}
\end{equation} 
\vspace{0.2cm}
\begin{equation} 
S_{g} = \frac{T_{mGPU}}{T_{GPU}}
\end{equation} 
\vspace{0.2cm}
\begin{figure}[h!]
\begin{center}
\includegraphics[scale=0.8]{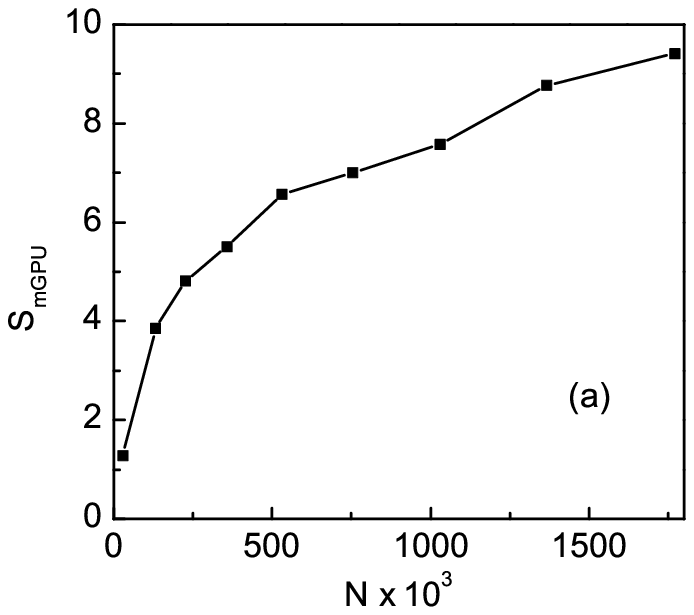}
\includegraphics[scale=0.8]{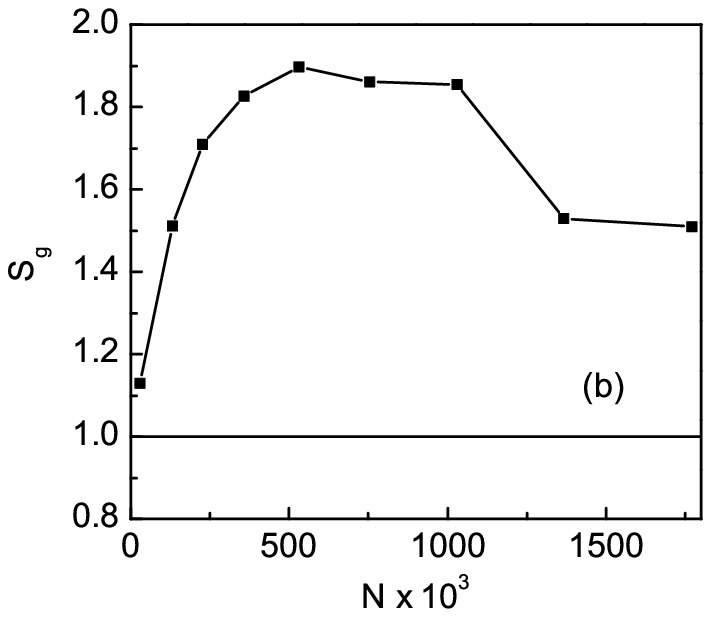}
\label{fig:subfigureExample}
\caption[Optional caption for list of figures]{Speedup of multiple-GPU implementation compared to (a) a single core and (b) a single GPU} 
\end{center}
\end{figure}

A maximum speedup of almost 10 compared to a single core and about 1.9 compared to a single GPU is observed. The combined computational capability of the two GPUs with a peak double precision floating point performance of over 1 Teraflops accelerated the computations several times. Compared to a single GPU, the overall speedup variation comprises clearly two different regions. In the first region, the increased computational power of the two GPUs contributes to the observation of an almost ideal speedup ($S_g=2$) compared to a single GPU. In the second region though, for DOF \textgreater $10^6$, the frequent communication and the larger amount of memory that has to be exchanged between the GPUs causes the GPU-to-GPU bus to saturate and as a result the communication to become a bottleneck that slows the computations down. In this implementation, the time required for synchronisation and data exchange of any kind (local and global communication) is measured and summed for the total communication time presented in Table \ref{mgpu-time}. For problems with DOF \textgreater 20000, the decreasing number of operations per GPU overcomes the delay caused by the communication, thus achieving a speedup greater than 1. According to Table \ref{mgpu-time}, as the problem size grows, the time required for the communication becomes less significant compared to the time required for computations. Nevertheless, for larger problems due to the higher processing power of the GPU compared to a core, the communication is still a major drawback, not allowing linear speedups to be achieved. 
\begin{table}[h!]
\begin{center}
\caption{Average time distribution on the GPUs versus problem size}
\begin{tabular}{lcc}
\hline\noalign{\smallskip}
DOF & Computation \%  & Total Communication \% \\
\noalign{\smallskip}\hline\noalign{\smallskip}
29791 & 79.84 & 20.16 \\
132651 & 83.45 & 16.53	 \\
226981 & 84.63 & 15.37  \\
357911 & 85.38 & 14.62  \\
531441 & 86.71 & 13.29  \\
753571 & 87.39 & 12.61  \\
1030301 & 87.95 & 12.05  \\
1367631 & 77.17 &	 22.93  \\
1771561 & 76.49 & 23.51 \\
\noalign{\smallskip}\hline
\end{tabular}
\label{mgpu-time}
\end{center}
\end{table}

\subsection{Comparison}

Table \ref{table2} shows the values of a factor defined as \emph{relative speed}:

\begin{equation}
\label{rel_s}
relative \ speed_{i} = \frac{T_{slowest}}{T_i}
\end{equation}
$T_{slowest}$ is the maximum wall clock execution time from all implementations for a specific problem size (i.e. number of DOF), $T_i$ is the minimum wall clock execution time for the implementation \emph{i} for the same DOF and \emph{i} is the implementation of the PGMRES (Cases A-D).

Apparently, the relative speed is equal to one for the slowest implementation at a specific number of DOF. As shown in Table 4, for small problems, PGMRES-classic on the shared memory architecture (Case B) is the fastest, while PGMRES-GPU (Case C) is the slowest. As the number of DOF increases, PGMRES-classic on the distributed memory architecture (Case A) becomes the slowest while PGMRES-mGPU becomes the fastest one. At DOF \textgreater $2\cdot 10^5$, comparable relative speeds of PGMRES-classic on Pegasus (distributed memory) and on one node of Andromeda (shared memory) are noticed, despite the Pegasus superiority in terms of speedup. This is due to the older hardware (CPUs, memory) of Pegasus compared to Andromeda's. Andromeda's single core performance is approximately 3.5 times faster than Pegasus. 

\begin{table}[h]
\begin{center}
\caption{Maximum relative speed. DM: Distributed Memory, SM: Shared Memory.}
\begin{tabular}{ccccc}
\hline\noalign{\smallskip}
DOF & PGMRES-classic (DM) & PGMRES-classic (SM) & PGMRES-GPU & PGMRES-mGPU  \\
\noalign{\smallskip}\hline\noalign{\smallskip}
29791 & 1.90 & 3.07 & 1.00 & 1.13  \\
132651 & 1.00 & 2.02	& 1.19	& 1.79 \\
226981 & 1.00 & 1.66 & 	1.15 & 	1.97 \\
357911 & 1.00 & 1.34	 & 1.27 &	2.33 \\
531441 & 1.00 & 	1.17	 & 1.27	 & 2.26 \\
753571 & 1.00 & 	1.32 & 	1.43 & 	2.37 \\
1030301 & 1.00 & 1.39 & 	1.73 & 	2.52 \\
1367631 & 1.00 &	 1.42 & 1.93 & 2.61 \\
1771561 & 1.00 & 1.51 & 2.01 & 2.83 \\
\noalign{\smallskip}\hline
\end{tabular}
\label{table2}
\end{center}
\end{table}

\section{Conclusions}
\label{conclusions}

In this work is presented an implementation of the preconditioned GMRES for large sparse linear systems ranging from a few thousands DOF to over one million on various parallel computing architectures. The partitioning technique used for the distribution of the computations to the various processes is briefly described and the basic aspects of the GPUs architecture are presented.

The main issue addressed in this work is the method that someone should choose in order to efficiently solve large sparse linear systems of equations depending on the hardware available. As far as one node of the distributed memory core/GPU cluster Andromeda is concerned, the shared main memory puts a serious limitation to the number of cores that can run concurrently, before memory access becomes a bottleneck. The use of a single GPU device can accelerate the computations several times. Using multiple GPUs even greater speedups are achieved. Furthermore, the use of multiple GPUs allows for the solution of even larger problems, because of the mesh partition. Nevertheless, those facts presuppose that the total GPU's main memory meets the memory requirements of the large-scale problem under study. Otherwise, the use of the distributed memory cluster Pegasus is preferred where the achieved speedup is not affected by the communication because of its fast network. Ongoing research involves a hybrid parallel implementation of the PGMRES capable to run on both multiple cores and GPUs.

\bibliography{./elsarticle-template-1-num.bib}

\begin{thebibliography}{10}

\bibitem{Smith}
B.~Smith, P.~Bjorstad, and W.~Gropp.
\newblock {\em Domain Decomposition. Parallel multilevel methods for elliptic
  partial differential equations}.
\newblock Cambridge University Press, 1996.

\bibitem{Saad_book}
Y.~Saad.
\newblock {\em Iterative methods for sparse linear systems}.
\newblock PWS Publishing Company, Boston, 1996.

\bibitem{Saad}
Y.~Saad and M.H. Schultz.
\newblock Gmres: A generalized minimal residual algorithm for solving
  nonsymmetric linear systems.
\newblock {\em SIAM J. Sci. Stat. Comp.}, \textbf{7}, (1986) 856-869.

\bibitem{papadra}
M.~Papadrakakis, G.~Stavroulakis, and A.~Karatarakis.
\newblock A new era in scientific computing, domain decomposition methods in
  hybrid cpu/gpu architectures.
\newblock {\em Comput. Methods Appl. Mech. Engrg.}, \textbf{200}, (2011)
  1490-1508.

\bibitem{ncuda}
https://developer.nvidia.com/content/cuda-10 (last~accessed 19.03.2013).

\bibitem{GMRES_distr}
H.E. Haiwu, C.~Bergere, and S.~Petiton.
\newblock A hybrid gmres/ls - arnoldi method to accelerate the parallel
  solution of linear systems.
\newblock {\em Comput. Math. Appl.}, \textbf{51}, (2006) 1647-1662.

\bibitem{Li}
Guangye Li.
\newblock A block variant of the gmres method on massively parallel processors.
\newblock {\em Parallel Comput.}, \textbf{23}, (1997) 1005-1019.

\bibitem{Erhel}
J.~Erhel.
\newblock A parallel gmres version for general sparse matrices.
\newblock {\em Electron. Trans. Numer. Anal.}, \textbf{3}, (1995) 160-176.

\bibitem{Pashos1}
G.~Pashos, M.E. Kavousanakis, A.N. Spyropoulos, J.A. Palyvos, and A.G.
  Boudouvis.
\newblock Simultaneous solution of large-scale linear systems and eigenvalue
  problems with a parallel gmres method.
\newblock {\em J. Comput. Appl. Math.}, \textbf{227}, (2009) 196-205.

\bibitem{Calvetti}
D.~Calvetti, J.~Petersen, and L.~Reichel.
\newblock A parallel implementation of the gmres method.
\newblock {\em Numerical Linear Algebra, eds. L. Reichel, A. Ruttan and R.S.
  Varga, de Gruyter}, (1993) 31-46.

\bibitem{Galoppo}
N.~Galoppo, N.K. Govindaraju, M.~Henson, and D.~Manocha.
\newblock Lu-gpu efficient algorithms for solving dense linear systems on
  graphics hardware.
\newblock {\em ACM/IEEE conference on Supercomputing}, (2005) 3.

\bibitem{Volkov}
V.~Volkov and J.~Demmel.
\newblock Lu, qr and cholesky factorizations using vector capabilities of gpus.
\newblock {\em Technical Report UCB/EECS-2008-49, EECS Department, University
  of California, Berkeley}, (2008).

\bibitem{Garland}
M.~Garland.
\newblock Sparse matrix computations on manycore gpu's.
\newblock {\em In Proceedings of the 45th annual Design Automation Conference
  (DAC '08)}, (2008) 2-6.

\bibitem{Boltz}
J.~Bolz, I.~Farmer, E.~Grinspun, and P.~Schroder.
\newblock Sparse matrix solvers on the gpu: conjugate gradients and multigrid.
\newblock {\em ACM SIGGRAPH 2003 Papers, San Diego, California, ACM, New York},
  \textbf{3}, (2003) 917-924.

\bibitem{sparse_GPU}
M.~Wang, H.~Klie, M.~Parashar, and H.~SudG.
\newblock Solving sparse linear systems on nvidia tesla gpus.
\newblock {\em Allen et al. (Eds.): ICCS 2009, Part I, LNCS 5544}, \textbf{3},
  (2009) 864-873.

\bibitem{Wang}
M.~Wang, H.~Klie, M.~Parashar, and H.~Sudan.
\newblock Solving sparse linear systems on nvidia tesla gpus.
\newblock {\em Allen et al. (Eds.): ICCS 2009, Part I, LNCS 5544}, (2009)
  864-873.

\bibitem{Moret}
I.~Moret.
\newblock A note on the superlinear convergence of gmres.
\newblock {\em SIAM J. Numer. Anal.}, \textbf{34}, (1997) 513-516.

\bibitem{Xian}
W.~Xian and A.~Takayuki.
\newblock Multi-gpu performance of incompressible flow computation by lattice
  boltzmann method on gpu cluster.
\newblock {\em Parallel. Comput.}, \textbf{37}, (2011) 521-535.

\bibitem{Ament}
M.~Ament, G.~Knittel, D.~Weiskopf, and W.~Straßer.
\newblock A parallel preconditioned conjugate gradient solver for the poisson
  problem on a multi-gpu platform.
\newblock {\em 18th Euromicro Conference on Parallel, Distributed and
  Network-based Processing}, (2010).

\bibitem{metis}
G.~Karypis and V.~Kumar.
\newblock A fast and highly quality multilevel scheme for partitioning
  irregular graphs.
\newblock {\em SIAM J. Sci. Comput.}, \textbf{20}, (1999) 359-392.

\bibitem{h_pegasus}
http://febui.chemeng.ntua.gr/pegasus.htm (last~accessed 19.03.2013).

\bibitem{h_andromeda}
http://febui.chemeng.ntua.gr/andromeda.htm (last~accessed 19.03.2013).

\bibitem{Pashos2}
G.~Pashos, E.D. Koronaki, A.N. Spyropoulos, and A.G. Boudouvis.
\newblock Accelerating an inexact newton/gmres scheme by subspace
  decomposition.
\newblock {\em Appl. Numer. Math.}, \textbf{60}, (2010) 397-410.

\bibitem{fem}
G.~Strang and G.J. Fix.
\newblock {\em An Analysis of the Finite Element Method}.
\newblock Englewood Cliffs, NJ, 1993.

\bibitem{mat_computations}
G.H. Golub and C.F. Van~Loan.
\newblock {\em Matrix Computations}.
\newblock The Johns Hopkins University Press, Baltimore, 1983.

\bibitem{Morgan}
R.B. Morgan.
\newblock A restarted gmres method augmented with eigenvectors.
\newblock {\em SIAM J. Matrix Anal. Appl.}, \textbf{16}, (1995) 1154-1171.

\bibitem{burrage}
K.~Burrage and J.~Erhel.
\newblock On the performance of various adaptive preconditioned gmres
  strategies.
\newblock {\em Numer. Linear Algebra Appl.}, \textbf{5}, (1998) 101-121.

\bibitem{ncuda2}
http://developer.nvidia.com/cuda-toolkit-40 (last~accessed 19.03.2013).

\end{thebibliography}
\bibliographystyle{unsrt}

\end{document}